\documentclass[usenatbib]{article}
\usepackage{graphicx}	
\usepackage{amsmath}	
\usepackage{amssymb}	
\usepackage{mathrsfs} 

\title{The status of isochrony in the formation and evolution of self-gravitating systems}
\author{Alicia Simon-Petit$^{(1)}$, J\'{e}r\^{o}me Perez,$^{(2)}$ and Guillaume Plum$^{(3)}$\\
    {\footnotesize (1,2) Applied Mathematics Laboratory, Ensta ParisTech, 828, Bd des Mar{\'e}chaux, 91120, Palaiseau, France }\\
	  {\footnotesize (3) GEPI, Paris Observatory, PSL University, CNRS, Place Jules Janssen 92195, Meudon, France}\\
	{\footnotesize (1) alicia.simon-petit@ensta-paristech.fr}\\
	{\footnotesize (2) jerome.perez@ensta-paristech.fr}
	}

\begin{document}
\maketitle

\begin{abstract}
In the potential theory, isochrony was introduced by Michel H\'{e}non in 1959 to characterize astrophysical observations of some globular clusters. Today, Michel H\'{e}non's isochrone potential is mainly used for his integrable property in numerical simulations, but is generally not really known. In a recent paper \cite{simonpetit:2017}, we have presented new fundamental and theoretical results about isochrony that have particular importance in self-gravitating dynamics and which are detailed in this paper. In particular, new characterization of the isochrone state has been proposed which are investigated in order to analyze the product of the fast relaxation of a self-gravitating system. The general paradigm consists in considering that this product is a lowered isothermal sphere (King Model). By a detailed numerical study we show that this paradigm fails when the isochrone model succeeds in reproducing the quasi-equilibrium state obtained just after fast relaxation.
\end{abstract}



\section{Introduction}

The formation and evolution of self-gravitating systems --- globular clusters or any kind of galaxies --- is a very complex physical process. 
It takes place on two distinct temporal scales:
\begin{itemize}
   \item A rapid phase on a few dynamical times of the system during which the systems collapse under its own gravitational field (\cite{BT08}). This fast time scale
            will be labelled as~$T_\mathrm{fa}\sim~T_\mathrm{dyn}$.
   \item A much longer phase, typically of the order of $T_\mathrm{lo}\sim \frac{N}{\ln N}T_\mathrm{fa}$ where $N$ is the number of particles in the system, during which the system
            slowly relaxes due to
            internal (two body relaxation cf.~\cite{heggie_hut_2003}, small potential fluctuations arising from their finite number of particles e.g. \cite{Heyvaerts}) 
            or external (tidal shocks, galactic potential, etc., e.g.~\cite{Gnedin:1997}) processes.
\end{itemize}
In addition to this time ordering, there also exists a spatial one which distinguishes between:
\begin{itemize}
   \item Singular systems, which fast formation process occurs in a single phase in a non-fluctuating external gravitational potential
            and which long-time evolution is not affected by merging. Typically, generic globular clusters and Low Surface Brightness (LSB) galaxies are singular systems.
   \item Composite systems, which formation process involves successive merging phases in a fluctuating gravitational field (e.g.~hierarchical scenario for High Surface Brightness (HSB) galaxies~(e.g.~\cite{1999MNRAS.303..188K}) or numerical simulations with clumps~(e.g.~\cite{vanAlbada82} for the initial study and~\cite{Trenti:2005,PhysRevE.71.016102} for more recent works)).
\end{itemize}
Using this nomenclature, a composite system can be viewed as the final result of a complex gravitational interaction between singular systems.
The evolution of a composite system over $T_\mathrm{lo}$ is a difficult problem and 
must be addressed in a cosmological framework. 
\\

In the present work, we are essentially interested in singular systems.
Their fast formation process results in a spherical density profile with specific properties: a large homogeneous core
containing half of the system mass and 
surrounded by a~$r^{-4}$ steep halo (\cite{RoyPerez}).
This initial characteristic profile is altered by the long dynamical evolution: the core shrinks
and the halo softens. The situation is rather different in composite systems.
As a matter of fact, on the first hand, 
modern simulations with clumps produce highly concentrated core-halo structures, see~\cite{Trenti:2005}.
On the other hand, the cosmological hierarchical scenario confers a cusp to HSBs' profiles.

The evolution of singular systems over
$T_\mathrm{lo}$ has been studied intensively, specially for globular cluster systems.
The relatively small sizes of such stellar clusters (about 1/1000 of the size of the host galaxy) 
make them stand as points in the galactic potential~(\cite{BT08}). Additionally, their typical dynamical time 
is about one tenth of that of their hosts. Therefore they can be considered as singular systems.


After decades, the lowered isothermal \cite{King} model  and its
generalizations (e.g. \cite{BT08} p. 307-311) appear to be quite universal to
describe any state of the evolution of globular clusters. Reasons for such a
success in the description of the dynamical properties of globular clusters
are physically understandable.
After the theory of violent relaxation by~\cite{LB67} one understood that this
full process leads to an isothermal sphere with an infinite mass (see e.g. \cite{BT08}
p.305-307). To get rid of this 
singularity, some physical improvements have been proposed.
These refinements
essentially consist in the introduction of some physical cut-off. When this
cut-off is crudely introduced by hands in the isothermal model, it produces the
original \cite{King} model. It is originally justified by spatial
limitation corresponding to the tidal cut-off imposed by the surrounding galaxy
in which the globular cluster evolves. When the cut-off is included as a
parameter of the statistical problem, it produces an isothermal sphere in
a box (ISB) which is a trademark problem of the gravitational statistical physics.
This fundamental problem has a centennial history and it finally appears to be
closely related with the King model (see \cite{Chavanis} and reference
therein). Studying the stability properties of this ISB in terms of the
conditions imposed on the box, theoretical astrophysicists were gradually able
to understand the dynamical history of globular clusters through the more and
more refined analysis of the so called gravothermal catastrophe (\cite{LBW},
\cite{Katz}, \cite{Patma}, \cite{Chavanis1}). The conclusion of this history
is now well established concerning the mass density (see for example a very
nice synthesis  in \S 7.5\ of \cite{BT08}): after the initial fast relaxation
process an isolated $N-$body system settles down in a spherical core-halo and
quasi-equilibrium state. Considering the sole density profile, this state appears well described by a lowered isothermal
sphere (King Model) and evolves adiabatically to a more concentrated lowered
isothermal state under the influence of slow relaxation. When the stability of
this lowered isothermal state -- governed by its density contrast -- is no more possible, the core of the system
shrinks. For example in our galaxy,  about 20\% of the globular
clusters possess the steep cusp in their surface-brightness profile (e.g.
\cite{Djorg1986})  that is predicted by models of post-collapse evolution. The
complementary set is not yet collapsed and possess a core-halo structure with a slope between $-4$ and $-2$ 
in the halo, fixed by the level of relaxation of the system. When the system is assumed to have a
non-constant mass function this mechanism is reinforced by mass segregation:
the most massive objects become concentrated in the high-density central core.

Since the "King-fit" scenario is globally accepted for the mass density (at
least for globular clusters in our Galaxy\footnote{For other hosts the situation is
not so clear, see e.g. \cite{McKey} for GC in the SMC.}), the interest of the
community progressively moved to the evolution of the globular cluster's mass function (see
\cite{Forbesetal} for the more recent review). This is a much more complicated and
tricky problem, which can be attacked only by observations or simulations.
This late interest let the theoretical approach of this problem at rest for a decade.

The three parameters of the King model essentially control the relative size
of its core and the slope of its halo. Monitoring the evolution of globular
clusters using King model seems to be a good idea, but several questions remain
posed. What are the physical characteristics of the initial steep core-halo $r^{-4}$
structure that is produced by the fast relaxation? In what sense is this initial
state of the slow relaxation a lowered isothermal sphere?

Quite recently, a fine analysis by \cite{Zocchi} points out that although King
models usually offer a good representation of the observed photometric
profiles, they often lead to less satisfactory fits to the kinematic profiles,
independently of the relaxation condition of the systems.
\\

In his pioneering paper Michel \cite{Henon58} proposed that
globular clusters can be isochrone. This proposition was based on the fact
that, in the late of the fifties, the observed globular clusters' mass density
distributions looked like the one of the isochrone model. The refinement of the
observations has actually revealed a wider diversity! First of all, the
mass density of globular clusters changes during their dynamical evolution.
Although they are characterized by a spherical core-halo structure almost all along
their life, the size of their constant density core and the slope of their
power law surrounding halo actually change during their evolution. When they
are dynamically young, observed or computed globular clusters are mainly
characterized by large cores surrounded by steep $r^{-4}$ halos. 

The old idea from Michel H\'{e}non to associate globular clusters to the isochrone model has then been progressively forgotten because of the King model's ability to reproduce their mass density distribution. It is probably why nobody has verified whether the result of the fast relaxation is truly isochrone, i.e. in a kinematic sense by checking orbital properties of its components. 

Excepted an interesting work by \cite{Voglis:1994} who remark a quasi-isochrone property in the result of very particular simulations of spherical cold collapse in a cosmological context, no work was directly addressed to test the isochrone status of some stage of the evolution of some kind of self-gravitating system.

Recently \cite{simonpetit:2017} reopened and widely reconsider the isochrony problem in the steps of Michel H\'{e}non.
In this fully detailed paper the essence of isochrony from mathematical concepts of group theory to the astrophysics of the generalized Kepler's Third Law,
passing through the new isochrone relativity theory. 
The aim of the present paper is two-fold. First, we want to summarize this huge paper in order to highlight its main results for practical uses in the context of galactic dynamics. Second, we propose
to implement some of these new concepts to address the old idea from Michel Hénon.
When translated in our nomenclature, this idea becomes: could the result of the fast relaxation of singular systems
be isochrone?

The paper is arranged as follows: in section~\ref{section2} we recall the main results of the general theory of isochrony recently obtained by~\cite{simonpetit:2017}. We detail in particular the generalized version of Kepler's third law which we will use as an isochrony test in the present paper. Section~\ref{sectionNumexp} is devoted to the numerical experiment which we propose in order to check the isochrone character of the product of fast relaxation. We then propose some analysis of the numerical results and conclude.

\section{Isochrony in 3D central potentials}
\label{section2}
Let us recall the definition of isochrony for spherically symmetric systems introduced in~\cite{Henon58} and briefly summarize the recent paper~\cite{simonpetit:2017}.
The latter is a complete and detailed paper about isochrony which contains a new approach to this problem and all proofs and details of some properties we use in this paper. In particular (i) it characterizes all continuous isochrone potentials, (ii) explains why the essence of isochrony is Keplerian, (iii) proposes the so-called isochrone special relativity theory and (iv) provides useful applications for our purpose like a generalization of the third Kepler law. 
 
Any test particle with a position-bounded and non-collapsing trajectory in a spherical self-gravitating system has its
radial distance $r(t) = \left|\mathbf{r(t)}\right|$ that oscillates between its value at pericenters and apocenters: $r_p=\min_t r(t)$ and $r_a=\max_t r(t)$.
Its radial distance is governed by the ordinary differential equation of motion
\begin{equation}
       \frac{1}{2} \left(  \frac{\text{d}r}{\text{d}t} \right)^2 + \frac{\Lambda^2}{2r^2} - \left( \xi -  \psi  \right) = 0  \label{fundedo}
 \end{equation}
where $\psi$ is the
radial gravitational potential associated to the system, $E=m\xi$ the energy of the test particle of mass $m$
and $L=m\Lambda$ the norm of its angular momentum.
The oscillation is characterized by 
its period
\begin{equation}
\tau(\xi,\Lambda)={2\int_{r_{p}}^{r_{a}}}\frac{dr}{\  \sqrt{2\left[  \xi-\psi \left(
r\right)  \right]  -\dfrac{\Lambda^{2}}{r^{2}}}} \label{radialperiod}%
\end{equation}
called radial period (see for example~\cite{BT08} p. 146).  A spherical system is isochrone if all its non-escaping and non-colliding particles admit radial periods that only depend on their energy, i.e. $\tau=\tau(\xi)$. This concept was first introduced by Michel H{\'e}non in a seminal paper in French (\cite{Henon58}, for an English translation see~\cite{Binney:2016}). 

In his paper, Michel H{\'e}non exhibited three isochrone central potentials. Completing his
analysis, there are actually four isochrone potential families. 
The evidence of the proof comes from the change of variables 
\begin{equation}
x=2 r^2 \ \text{ and } \ Y(x) = x \psi(x)
\end{equation}
exploited by Michel H{\'e}non. In these variables, $Y$ is related to the potential $\psi$. One 
can further show that $\psi$ is isochrone if, and only if, the graph of $Y$ is a parabola.
This parabola property geometrically characterizes an isochrone system. If the origin
of the plane $(x,y)$ is defined such that it coincides with the center of symmetry 
of the spherical system ($r=0$), then the parabola intersects the vertical
axis $(Oy)$ at least once. Increasing potentials are associated to positive mass particles and
are contained in the convex branches of the parabolas, whereas the concave branches contain
decreasing potentials (in parabolas with non-vertical symmetry axes).

A reduction of the isochrone potential search is possible through the action of geometrical
transformations. In Michel H{\'e}non's variables, the radial equation of motion \eqref{fundedo} writes
\begin{equation}
\displaystyle \tfrac{1}{16}\left(  \frac{dx}{dt}\right)  ^{2}+\Lambda^{2}%
=x\xi-Y\left(  x\right)  \text{.} \label{fundamentalode}%
\end{equation}
In this equation, $\xi$ and $\Lambda$ are the two free parameters of the problem. The whole set of isochrones is obtained spanning the allowed $(\xi,\Lambda)$ space when the graph of $Y(x)$ is a parabola in the $(x,y)$ plane. This span is possible with two simple transformations: (i) adding a constant to $Y$ which corresponds to adding a constant to the angular momentum
or (ii)  adding a linear term in $x$  which corresponds to adding a constant
to the energy. Both these two transformations do not affect the type
of the potential described by $Y$. In terms of parabolas, the first transformation named 
$J_{0,\lambda}: (x,y) \mapsto (x,y+\lambda)$ is simply a vertical translation. The second transformation
is a transvection $J_{\epsilon,0}: (x,y) \mapsto (x,y+\epsilon x)$. It swivels the parabolas by
inclining its symmetry axis, with a slope between $-\infty$ and $+\infty$, and preserving
the intersection of the parabola with the ordinate axis, as well as the abscissa of its vertical tangent,
when it exists. The two transformations generate a subgroup 
\begin{equation}
\mathbb{A} = \{   J_{\epsilon,\lambda}; \, (\epsilon,\lambda) \in \mathbb{R}^2  \} 
\end{equation}
of the
affine group and reduces the search of isochrone potentials to four classes of parabolas and thus potentials.
Each class is generated by one of the four following parabolas 
under the action of $\mathbb{A}$:
\begin{itemize}
   \item the parabola with a vertical tangent at the origin $\left(y\propto\pm\sqrt{x}\right)$; it is associated to a Kepler potential,
            $\psi_{\mathrm{ke}}\left(  r\right)  =-\dfrac{\mu}{r}$, 
            with $\mu>0$ related to the central mass;
   \item the classic parabola ($y\propto x^2$); it is associated to a harmonic potential, 
            $\psi_{\mathrm{ha} }\left(  r\right)  =\frac{1}{2}\omega^{2}r^{2}$, with $\omega$ real;
   \item a parabola that intersects $(Oy)$ twice:
             \begin{itemize}
                \item if it opens to the right, it is associated to a H{\'e}non potential,
                          $\psi_{\mathrm{he}}\left(  r\right)  =-\dfrac{\mu}{b+\sqrt{b^{2}+r^{2}}}$ where $b>0$;
                \item if it opens to the left, it is associated to an isochrone potential, called 
                          bounded or pseudo-H{\'e}non potential, $\psi_{\mathrm{bo}}(r)=\dfrac{\mu%
                           }{b+\sqrt{b^{2}-r^{2}}}$ for $r\in[0,b]$ where $b>0$.
             \end{itemize}
\end{itemize}
The physical properties of these potentials are discussed in~\cite{simonpetit:2017}.

 Therefore, the action of the affine subgroup $\mathbb{A}$ divides the isochrone
potentials, or equivalently the associated parabolas, into four
families: Kepler, harmonic, H{\'e}non and bounded (potentials). 
\\

The previous geometrical approach
classifies and completes the set of continuous
isochrone potentials. However, it splits them into four disjoint families and does not help
understanding the nature of isochrony. As a matter of fact, there exists an intrinsic link
between all isochrone systems as shown below.

Using Michel H{\'e}non's variables, a radially periodic orbit 
$r_{0}(t_{0})$ in a potential $\psi_0$ can be designed by $(\xi_0 x_0,Y_0)$ and~\eqref{fundamentalode} with its integral of motion $(\xi_0,\Lambda_0)$.
We consider its image when transforming the affine coordinate system $(w_1=\xi x,w_2=y)$ by 
$\mathbf {w_0 } = \left[\xi_0 x_0,y_0\right]^\top \mapsto \mathbf {w_1 } = \left[\xi_1 x_1,y_1\right]^\top$.
Assume $\Lambda_0=\Lambda_1=\Lambda$, i.e. the radial orbits share the same area law, 
and $\xi x - Y $ is conserved by the transformation as initiated by~\cite{LyndenBell}. 
If $Y_0$ is isochrone, then
$ Y_1$  is also isochrone if, and only if, the transformation is linear: 
$\mathbf{w}_{1}=B_{\alpha,\beta}\left(  \mathbf{w}_{0}\right)
$ with 
\begin{equation}
B_{\alpha,\beta}=\left[
\begin{array}
[c]{cc}%
\alpha & \beta \\
\alpha-1 & \beta+1
\end{array}
\right]  ,\  \left(  \alpha,\beta \right)  \in \mathbb{R}^{2}.
\end{equation}
This transformation generalizes the so-called Bohlin or Levi-Civita transformation\footnote{In fact this transformation was initiated by Newton in his Principia, formalized by \cite{goursat} and \cite{Darboux} followed by \cite{bohlin} and finally \cite{levicivita}.}, linking all isochrone potentials and orbits together.
When $B_{\alpha,\beta}$ is symmetric, it is named \textit{i}Bolst, for symmetr\underline{\textbf{i}}c \underline{\textbf{bo}}h\underline{\textbf{l}}in boo\underline{\textbf{st}},  and is written $B_\gamma$ where $\gamma=\alpha+\beta\neq0$. It is essentially the composition
of a homothety with a hyperbolic rotation. The length of vectors is then modified by
\begin{equation}
\left(\xi_1 x_1\right)^2 - y_1^2=\gamma \left[\left(\xi_0x_0\right)^2-y_0^2\right] . 
\end{equation}
Under the \textit{i}Bolst group action, the Keplerian potentials generate the isochrone potentials set. More generally, \textit{i}Bolsts provide a characterization of isochrone systems and orbits:
an orbit is isochrone if, and only if, it is the \textit{i}Bolsted image of a Keplerian orbit.

 Considering the canonical axis $Ox=\mathbb{R}\mathbf{i}$ and $Oy=\mathbb{R}\mathbf{j}$, a natural reference frame $(O,\mathbf{u},\mathbf{v})$ is attached to each isochrone parabola. These two natural vectors respectively generate the symmetry axis of the parabola for $\mathbf{u}$ and the tangent to the parabola at the origin $O$ for $\mathbf{v}$. For the Keplerian parabola we have $\mathbf{u}=\mathbf{i}$ and $\mathbf{v}=\mathbf{j}$, for the harmonic there is an inversion and one have $\mathbf{u}=\mathbf{j}$ and $\mathbf{v}=\mathbf{i}$. This inversion corresponds to the Bohlin--Levi-Civita transformation.
Remarkably, it is always possible to map the reference frame of any isochrone parabola to its Keplerian
equivalent with an \textit{i}Bolst,
such that $(\mathbf{u},\mathbf{v}) = B_\gamma(\mathbf{i},\mathbf{j})$, 
see~\cite{simonpetit:2017} for the general case.
Consequently, isochrone orbits in $(O,\mathbf{i},\mathbf{j})$ are Keplerian in their reference frames 
$(O,\mathbf{u},\mathbf{v})$. 
Hence isochrony appears as a relative Keplerian property: any isochrone system is Keplerian when seen in its reference frame,
this is isochrone relativity~\cite{simonpetit:2017}.

The parallel with special relativity provides a direct comprehension
of isochrone relativity. The isochrone law of dynamics writes in
the same way in all reference frames defined by the \textit{i}Bolsts.
The "\textit{isochrone interval}" $\xi x-Y$ is conserved and the
energy and potential terms are linearly combined. 
As in special relativity where the nature of time, light
and space-like vectors 
is conserved by Lorentz transformations, particular-like cones are preserved by 
\textit{i}Bolsts and 
characterize aperiodic, radial and radially periodic orbits in the isochrone relativity. The correspondance between 
radially periodic orbits through the \textit{i}Bolst relies on an appropriate choice
of affine coordinates system. For any isochrone radially periodic orbit, one can then 
easily construct its Keplerian or harmonic associated 
periodic orbit, even though the time
does not run identically for the correspondant particles.
\\

Eventually, one of the most noticeable property of Keplerian systems is the
Kepler Third law, that proportionally relates the squared radial period $\tau^2$ of any particle to the cube of its semi-major axis $a^3$. The isochrone relativity enables one to understand that this relation holds for any isochrone potentials
after adapting the lengths. 

 As in the Kepler potential, characteristic semi-major axes $a$ are defined in any isochrone potential by: 
\begin{enumerate}
\item 
$a=\frac{1}{2}\left( r_{a}+r_{p}\right)$ in a Kepler potential;

\item 
$
a=\left( \frac{1}{2}\right) ^{2/3}R$ in a homogeneous bowl of radius $R$;

\item
$
a=\frac{1}{2}\left( \sqrt{b^{2}+r_{a}^{2}}+\sqrt{b^{2}+r_{p}^{2}}
\right)$  in a H\'{e}non potential;

\item 
$
a=\frac{1}{2}\left( \sqrt{b^{2}-r_{a}^{2}}+\sqrt{b^{2}-r_{p}^{2}}
\right) 
$ in a bounded potential.
\end{enumerate}
The Kepler Third Law is then generalized for any isochrone system:
\begin{equation}\label{eq:KeplerThirdLawSMA}
\tau^2 = \frac{4\pi^2}{\mu} a^3,
\end{equation}
where $\mu$ is the potential parameter which appears in the definition of $\psi_{\mathrm{ke}}$, $\psi_{\mathrm{bo}}$ or $\psi_{\mathrm{he}}$, and $\mu = GM = \omega^2 R^3$ for the homogeneous bowl. This isochrone orbital property holds for each orbits, it is then a key microscopic criteria for the isochrone analysis of systems of given macroscopic density profile.

\section{Numerical experiments}
\label{sectionNumexp}
The objective of this section is twofold. 
Fisrt, we want to check the old idea from Michel H\'{e}non which is in our nomenclature : do the fast collapse of a singular system is an isochrone system. Second, we want to check if the relevant King model commonly adopted to describe the result of this fast collapse passes or not the isochrone test.
\subsection{Description of the simulations}
We have performed two accurate $N-$body simulations using the latest version
of the \texttt{Gadget-2} code \cite{Springel2005}. Only the treecode part of
treePM\ algorithm is employed. These simulations contain $N=3.10^{4}$ equal
mass $m$ particles in a typical radius of the order of unity in the units
of the simulations. These units are such that the gravitational constant
$G=1$ and the total mass $M=Nm=1$. As shown by \cite{RoyPerez}, this number of particles is sufficient for
our purpose of capturing the physics of the problem with reasonable runtimes. These physical self-gravitating systems we have computed are a H\'{e}non sphere and a King model.

The H\'{e}non sphere (e.g. \cite{Henon64}) is particularly suited for our purpose. On the first hand it is clearly the simpliest idealisation of what we have called an singular system. On the second hand, it is known
to preserve well its spherical nature during the course of the dynamics even
being simulated with a $N-$body technique (see for example \cite{RoyPerez}). In this model the initial
phase-space distribution function is isotropic, spatially homogeneous, gaussian distributed in
velocity space and given by%
\begin{equation}
f_{H}\left(  r,v,j\right)  =\left \{
\begin{array}
[c]{lc}%
\rho_{0}\left(  2\pi \sigma_{v}^{2}\right)  ^{-3/2}
\exp \left(-\frac{1}{2}\frac{v^{2}+j^{2}/r^{2}}{\sigma_{v}^{2}}\right)  , & \text{ for }r<R;\\
0 & \text{ if }r>R.
\end{array}
\right. 
\end{equation}
Considering $E_k=\sum_{i=1}^{N}\frac{1}{2}m\mathbf{v}_{i}^{2}$ and $E_p=\sum_{i=1,j>i}^{N}-\frac{m^2}{\left|\mathbf{x}_{i}-\mathbf{x}_{j}\right|}$ respectively the total kinetic and potential energy of the system, when the initial virial ratio $\kappa=2E_k/E_p$ is in the interval $\left]-1,0\right[$  the system collapses to an equilibrium state in a few dynamical times (e.g. \cite{RoyPerez}). Provided that the initial state is not too cold ($\kappa\in\left]-1,-0.25\right[$) the radial orbit instability (see \cite{MarechalPerez} for a review of this fundamental process) does not occur and the equilibrium state is spherical. Its mass density is characterized by a core-halo structure. It is very well established that the core of this structure contains roughly half of the mass of the system and the halo is well approximated by a power law: $\rho(r)\propto r^{-4}$ (see e.g. \cite{RoyPerez} or \cite{Joyceetal}\footnote{The figure 14 of this paper is particularly explicit about this result}). For our purpose we have studied a H\'{e}non sphere with $\kappa=-0.5$, an initial size $R=2$ which gives after collapse a typical size  $R_{50}=1$ in the units of the simulation.

The King model is a stable spherical isotropic equilibrium state of the Vlasov-Poisson equation. 
It means that if there is no relaxation its distribution function does not change in time. This distribution function is given by 
\begin{equation}
f_{K}\left(  E\right)  =\left \{
\begin{array}[c]{lc}%
\rho_{0}\left(  2\pi \sigma_{\epsilon}^{2}\right)  ^{-3/2}
\exp \left(-\frac{E_{\ell}-E}{\sigma_{\epsilon}^{2}}\right) -1 , & \text{ for
}E<E_{\ell};\\
0 & \text{ if }E>E_{\ell}.
\end{array}
\right. 
\end{equation} 
The mean field potential $\psi(r)$ associated to this distribution function is given by the Poisson equation
\begin{equation}
\Delta \phi(r) = 4\pi m G \rho_0
\left\{
\sqrt{\frac{4\phi}{\pi\sigma_{\epsilon}^2}}\left( 1+\frac{2\phi}{3\sigma_{\epsilon}^2}\right)+
\mathrm{e}^{\frac{\phi}{\sigma_{\epsilon}^2}}\mathrm{erf}\left(\sqrt{\phi}/\sigma_{\epsilon}\right)
\right\}
\end{equation}
where 
\begin{equation}
\phi(r)=E_{\ell}-m\psi(r)\;\mathrm{and}\;\mathrm{erf}(x)=\frac{2}{\sqrt{\pi}}\int_0^x \mathrm{e}^{-u^2}du\,.
\end{equation}
The King model has three free parameters which are: 
\begin{itemize}
\item the liberation energy $E_{\ell}$ which is the cutoff in the energy space introduced to cure the infinite mass problem of the isothermal model;
\item the depth of the potential well $\psi(0)$ at the origin;
\item the energy variance $\sigma_{\epsilon}^2$.
\end{itemize}
These parameters are gathered in one by introducing $W_{0}=\frac{\phi(0)}{\sigma_{\epsilon}^2}>0$. As said before, King models are core-halo structures (see for example \cite{BT08} p.308-309): the size of the core is a non trivial function of $r_c=\sigma_{\epsilon}(4\pi mG\rho_0)^{-1/2}$ while $W_0$ controls the steepness of the power law $r^{\alpha}$ in the halo of the mass density profile. The slope $\alpha$ varies from $\alpha\simeq-2$ (the isothermal value) when $W_0$ is large, typically greater that 12, to $\alpha\leq-5$ when $W_0$ becomes smaller than 3. The large-energy cut-off of the King model introduces a sharp cut-off in the mass density profile for large values of the radius. The slope of the halo we are speaking about concerns the region surrounding the core where the mass density is not less than typically $\rho_{0}/10^3$. This value corresponds to observable and then measurable values of the projected luminosities profiles.  For our purpose we have studied a King model with $W_{0}=9$. This value corresponds to the best-fit concentration parameter $c=2$ proposed in figure 1 of \cite{Djorg1986} thanks to the correspondence figure 4.9 p. 310 in \cite{BT08}. The typical size of our King model is $R_{50}\simeq1$.

The parameters for \texttt{Gadget} runs were adapted to our purpose:
\begin{itemize}
\item The softening length of the gravitational force is set to $\varepsilon
=\left(  \frac{4\pi}{3N}\right)  ^{1/3}R_{50}$ with $R_{50}\simeq1$ for both simulations. This value corresponds to an
estimation of the initial mean interparticle distance, it is a bit larger than the usual value for $\varepsilon$ in standard simulations. Nevertheless, it is well adapted for
our purpose for which we want to minimize the effect of two body relaxation in
order to study the properties of orbits in a frozen collisionless equilibrium gravitational
potential. The obtained value is adequate for the softening of the force, according to the criterion proposed by~\cite{2000MNRAS.314..475A}.
This value is also sufficient to solve the collapse problem when
the initial virial ratio of the H\'{e}non sphere is not too small and the corresponding collapse is not too fast e.g.
$\kappa=-0.5$ (see \cite{RoyPerez}). 

\item In \texttt{Gadget}, each particle has its individual time step bounded
by $\delta t=\min \left[  \  \delta t_{\max},\sqrt{2\eta \varepsilon/\left \vert
\mathbf{a}\right \vert }\right]  $, where $\mathbf{a}$ is the acceleration of
the particle and $\eta$ is a control parameter. We choose $\eta=0.025$ and
$\delta t_{\max}=0.01$. This ensures an energy conservation of the order of 1\%
for each run.

\item The tolerance parameter controlling the accuracy of the relative
cell-opening criterion (parameter designed by \texttt{ErrTolForceAcc} in the
documentation of \texttt{Gadget}, see equation 18 of \cite{Springel2005}) is
set at $\alpha_{F}=0.005$.
\end{itemize}

The duration of each simulation is 300 equilibrium dynamical times for the H\'{e}non sphere and 500 equilibrium dynamical times for the King model.

\subsection{Orbit monitoring and simulation results}
\begin{figure}
\includegraphics[scale=0.6]{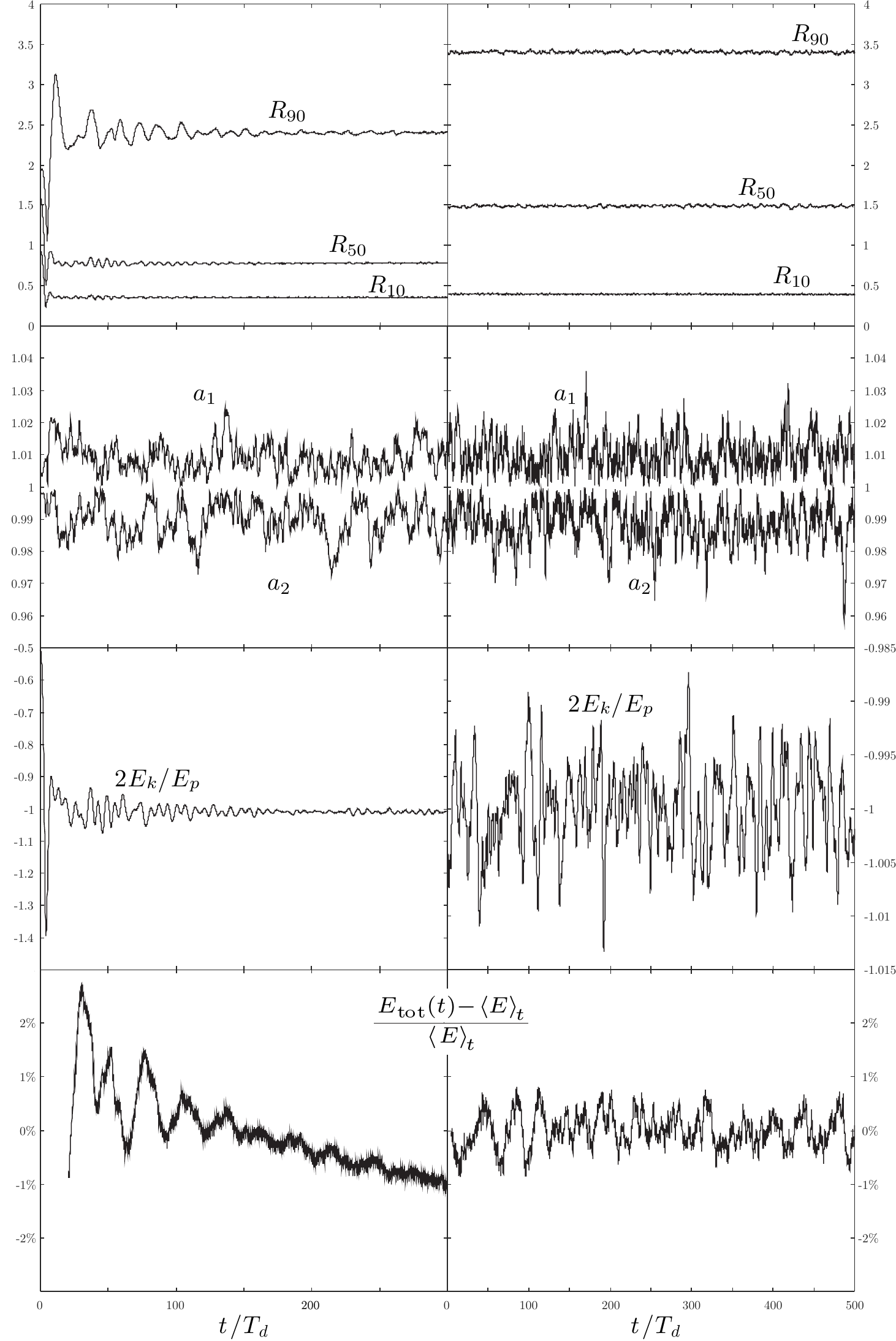}
\caption{Evolution of the mean physical characteristics of the simulation: on the left side, the H\'{e}non sphere with $\kappa=-0.5$ at $t=0$; on the right side the King model with $W_{0}=9$ at $t=0$. Notice the different time scales and the adapted $y-$values for each panel.}
\label{analyse:simus}
\end{figure}

The evolution of the physical parameters of the simulations are presented on figure \ref{analyse:simus} the left side concerns the H\'{e}non sphere and the right one the King model: 
\begin{itemize}
\item
$R_{90}$, $R_{50}$ and $R_{10}$ respectively represent the radii containing 90\%, 50\% and 10\% of the total mass of the system. They are plotted in the top two panels. The King model is a stable equilibrium state, so these three quantities remain constant during the dynamical evolution. The initial state of the H\'{e}non sphere with $\kappa=-0.5$ suffers Jeans' instability and collapses to a steady state in a few dynamical times. The typical size of each system stays of the order of unity.
\item
Computing the eigenvalues $\lambda_1>\lambda_2>\lambda_3$ of the inertia matrix of the system, the quantities $a_1=\lambda_1/\lambda_2$ and $a_2=\lambda_3/\lambda_2$ are usually called the axial ratios of the system. Both of these quantities are equal to $+1$ when the system is a sphere. We can see on the next two panels that our simulations remain spherical all over their dynamical evolution.
\item
The virial ratio $\kappa$ is equal to $-1$ when the system is at equilibrium. We can see on the third line panels that the King sphere remains at equilibrium and that the H\'{e}non sphere quickly joins such a state just after a fast relaxation initial phase.
\item
For such hamiltonian systems, the total energy $E_{tot}=E_p+E_k$ is conserved during the time evolution. We observe that in the King model this conservation is properly respected in a mean sense. For the H\'{e}non sphere, this is the same after the warm collapse of the system.
\end{itemize}

\begin{figure}
\includegraphics[scale=0.8]{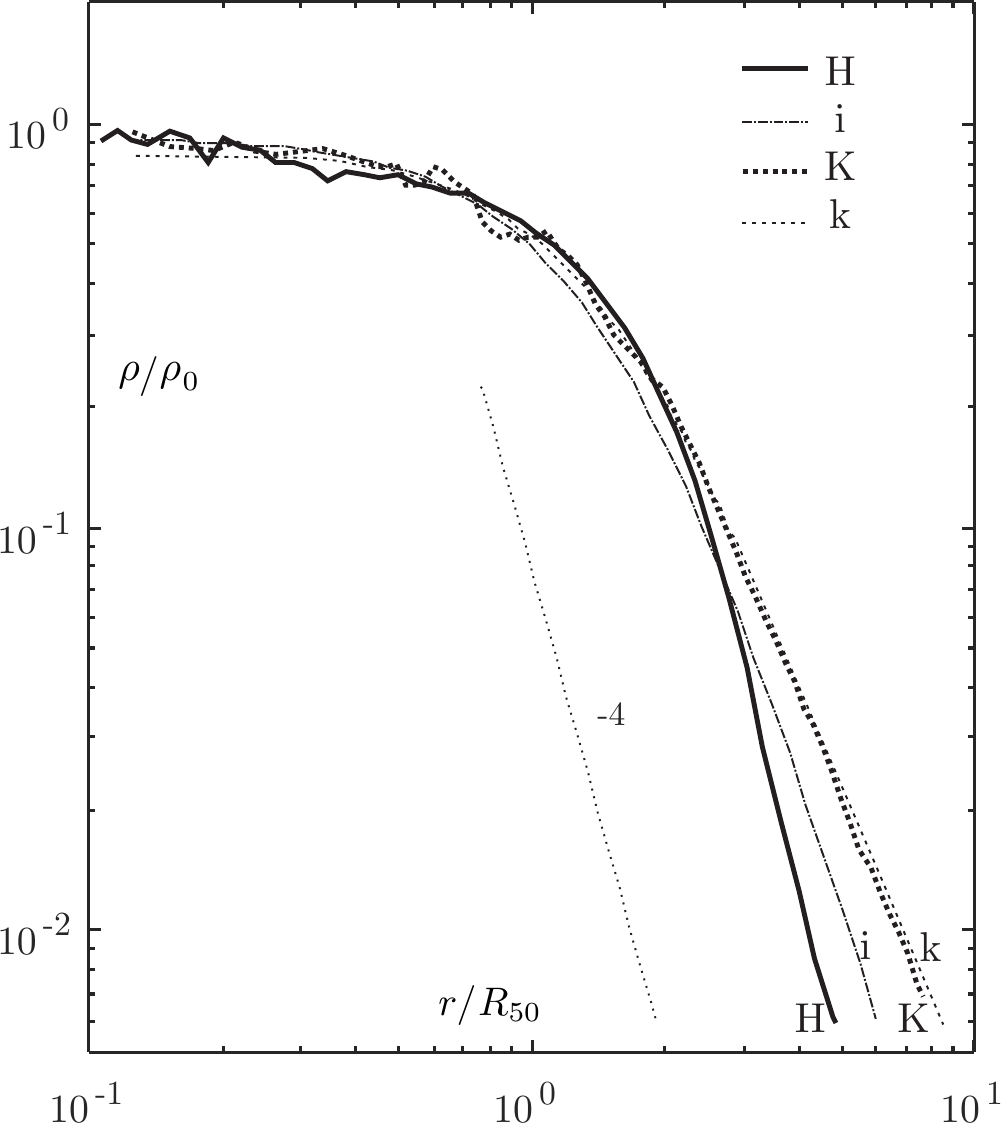}
\caption{Mass density comparison: H represents the H\'{e}non sphere at $t=100 T_d$, K the computed King model at $t=200 T_d$, i and k are respectively the plots of the isochrone($b=0.36$) and the King($W_0=9$) normalized theoretical mass densities.}
\label{densities}
\end{figure}

Counting particles in concentric spherical shells, we have computed the radial mass density for our simulation every 10 dynamical times. As expected the mass density of the King sphere does not evolve at all over the 500 dynamical times computed. As expected too, the H\'{e}non sphere initially collapses in a few dynamical times and reaches a steady state characterized by a core-halo structure: the size of the constant mass density core is roughly the radius containing half of the total mass of the system; the slope of the power-law like surrounding halo is roughly -4. Once it is formed ($t\sim 10T_d$), this core-halo structure does not evolve at all until the end of our computation at $t=300\;T_d$. In figure \ref{densities}, we have plotted these mass densities on the same graph in order to compare them. A rough analysis does not reveal differences between the results of our warm collapse (H) and simulated King model with $W_0=9$ (K). In addition we note that we can adjust the unique parameter $b$ of the H\'{e}non potential $\psi_{\mathrm{he}}$ defined in section \ref{section2} to get an isochrone mass density (i) which resembles to the one of (K) or (H).  In fact the plots of figure \ref{densities} clearly show that the density analysis is not conclusive concerning the nature of the equilibrium we obtain after the fast relaxation: it could be fitted by either a King model or an isochrone. Hence, a precise kinetic analysis is needed to reach firm conclusions. 
The kinematic analysis we propose consists in a fine study of orbits. 
In each simulation a subset of $n=200$ randomly chosen particles was monitored: three
quantities are archived at each time step $t$, namely the position
$\mathbf{r}(t)$, the velocity $\mathbf{v}(t)$ and the gravitational potential
$\psi \left(\mathbf{r},t\right)$ imposed by the $N$ bodies of the system at
the position $\mathbf{r}$ at time $t$. We point out that we have computed this
potential using the Gadget-2 Treecode algorithm, hence with the same softening parameter
$\varepsilon$ defined above. Analyzing these quantities, we can determine, for each monitored particle and when it exists, the period $\tau$ of its radial distance and in all cases its total energy $e(t)=\frac{1}{2}m\mathbf{v}^2(t)+m\psi \left(\mathbf{r},t\right)$. 

\begin{figure}
\includegraphics[width=\linewidth]{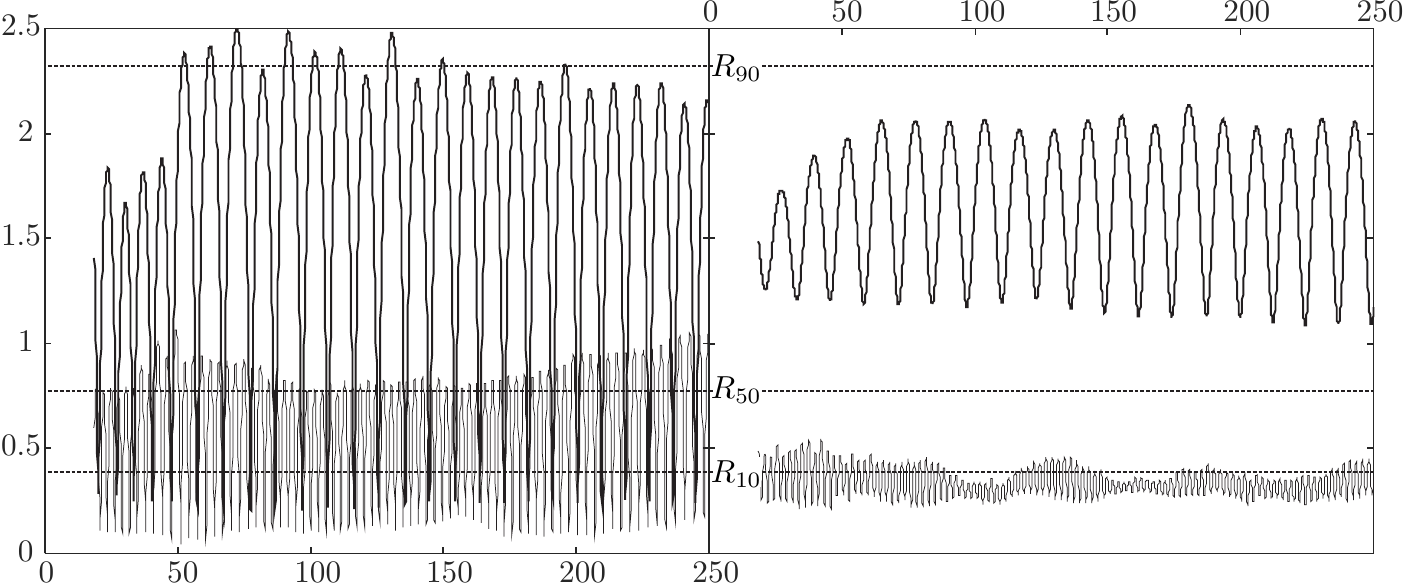}
\caption{Orbits of 4 particles extracted from the H simulation. On $y-$axes, the dotted lines indicate the typical lengths $R_{90}$, 
$R_{50}$ and $R_{10}$ of the whole system. The time is indicated in the simulation's $T_d-$units on the $x-$axes.}
\label{orbits}
\end{figure}

\begin{figure}
\centerline{\includegraphics[width=.9\linewidth]{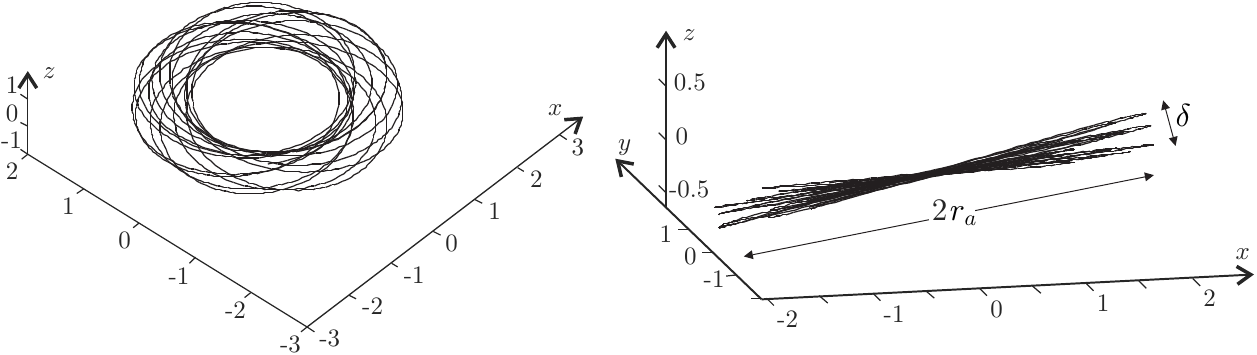}}
\caption{3D representation of an orbit extracted from the K simulation: A view from above is presented in the left panel while the side-on view is presented in in the right one.}
\label{orbitplan}
\end{figure}

The analysis of the period of particles is a tricky job. A priori, we deal with orbits of negative energy particles in a spherical self-gravitating system, and thus in a central potential. In such conditions, the orbit is planar and the radial distance $r(t)=\left|\mathbf{r}(t)\right|$ is a periodic function of time, i.e. there exists $\tau\in\mathbb{R}^+$ s.t. $r(t+\tau)=r(t)$. To find this $\tau$ from numerical data and then the values of $r_a$ and $r_p$, the simplest way should consist in the use of the FFT algorithm; but as it is could be seen on figure \ref{orbits}, in a practical way this method is not precise as it should be expected for several reasons.  
Depending on the particle properties, its orbit could be concentrated in the deep center of the system (see the lower orbit on the right side of Fig. \ref{orbits}), be spanning only the halo (see the upper orbit on the right side of Fig. \ref{orbits}), be spanning all the system or all its core (see the left side of Fig. \ref{orbits}) or be even more special. When the particle experiences the deep core, the two body effects could influence its dynamics; although the radial distance is periodic this function is modulated both in phase and in amplitude. The phase modulation due to the high density values is weaker when the orbit stays in the outskirts of the system where only large scale oscillations of the potential modulate its amplitude. These effects affect  both K and H simulations and introduce various and uncontrollable biases when we compute the period using FFT on the whole data. Instead, in order to get the right value of the period with the smallest significative error, we carry out a hand-made analysis of each orbit. We first check the planar property of the orbit: we determine the mean angular momentum, compute the orthogonal plane to this mean vector and reject orbits with an amplitude of azimuthal oscillation $\delta$  (see figure \ref{orbitplan}) around this plane less than $20^{\circ}$, such that $\delta \geq 0.2 r_a$. 
When the orbit is planar, we determine 
the coordinates $(r_{a,i},t_{a,i})_{i=1,\cdots,\mathcal{N}}$ of $\mathcal{N}$ successive maxima of the function $r(t)$. The value of the integer $\mathcal{N}$ depends on each orbit considered. We set $\mathcal{N}\geq\mathcal{N}_{\min}=5$ in order to compute the period at least over $\mathcal{N}_{\min}$ oscillations. This minimal value allows us to initiate the computation of the period 
$\tau=\frac{t_{a,\mathcal{N}}-t_{a,1}}{\mathcal{N}-1}$, the value of $\mathcal{N}$ is incremented while $\sigma_{\tau}/\tau<5\%$ where $\sigma_{\tau}$ is the standard deviation of the sequence of instantaneous periods $\tau_i=t_{a,i+1}-t_{a,i}$ for $i=1,\cdots,\mathcal{N}-1$. When this algorithm does not converge the orbit is rejected as not periodic. When it gives a unique value this result is compared with the data. When several values of the period are possible during several phases of the orbit, it is rejected as multiperiodic\footnote{By \emph{several} we mean more dispersed than $5\%$ around the mean value. This possibility occurs when there is a strong two body interaction in which the characteristics of the orbit (period, apocenter, pericenter) are modified}. When the period of the orbit is confirmed we determine its apocenter $r_a=\frac{1}{\mathcal{N}}\sum_{i=1}^{\mathcal{N}}r_{a,i}$. Computing the sequence of the minima of $r(t)$ in the interval $\left[t_{a,1},t_{a,\mathcal{N}}\right]$ we obtain the pericenter of the orbit in the same mean sense. Computing the mean value of the energy $E=\left\langle e\right\rangle _t$ on this same time interval, we get the mean energy of the particle. The standard deviations of $e$, $\tau$, $r_a$, and $r_p$ are used for the uncertainty analysis: the amplitude of error bars in the plots are twice the standard deviation.

Using this manufactured but precise algorithm, we are able to extract with a good level of confidence the energy, period, apocenter and pericenter of $n_{\mathrm{H}}=155$ orbits for the H simulation and $n_{\mathrm{K}}=172$ orbits for the K simulation among the $n=200$ monitored for each one.

\begin{figure}
\centerline{\includegraphics[width=\linewidth]{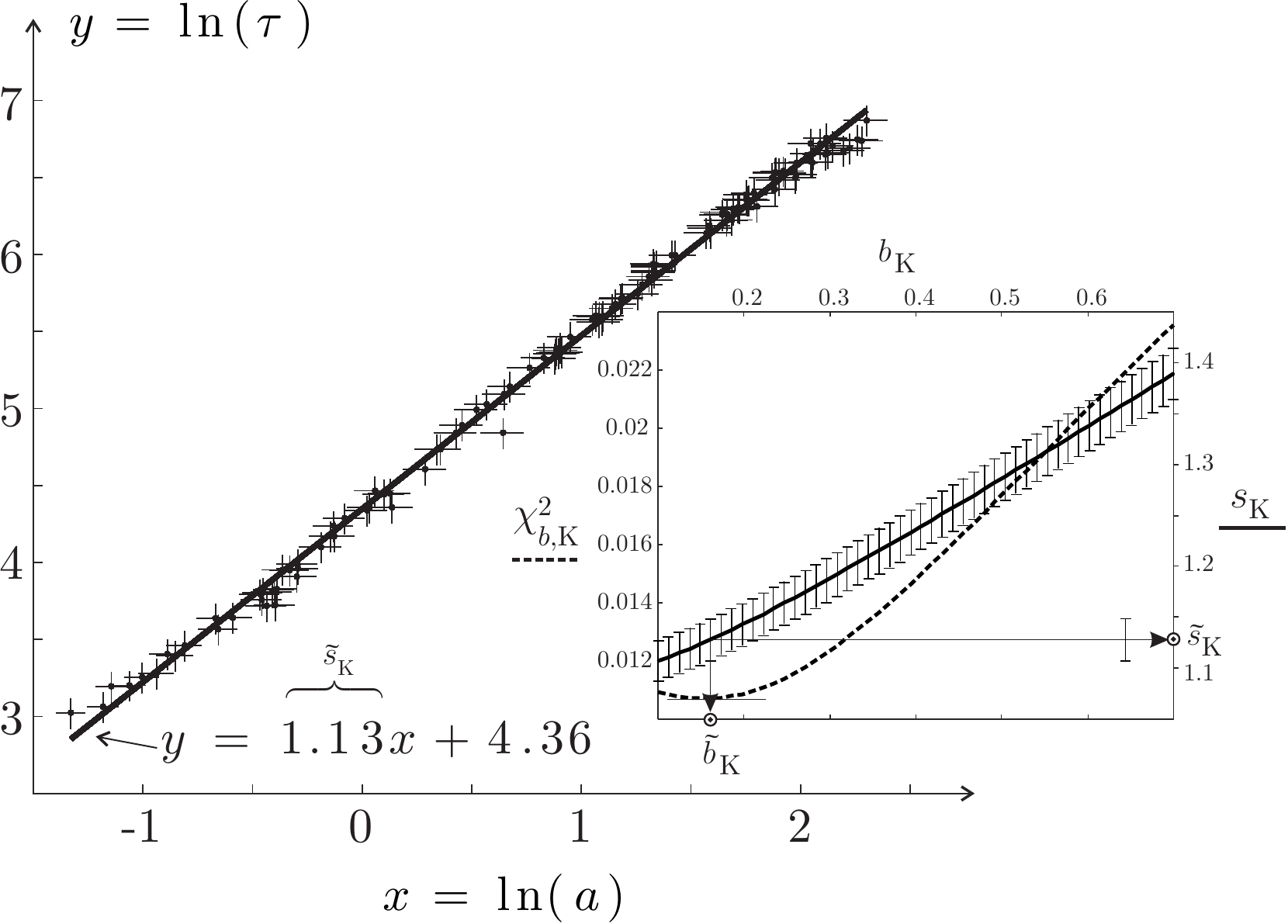}}

\centerline{\includegraphics[width=\linewidth]{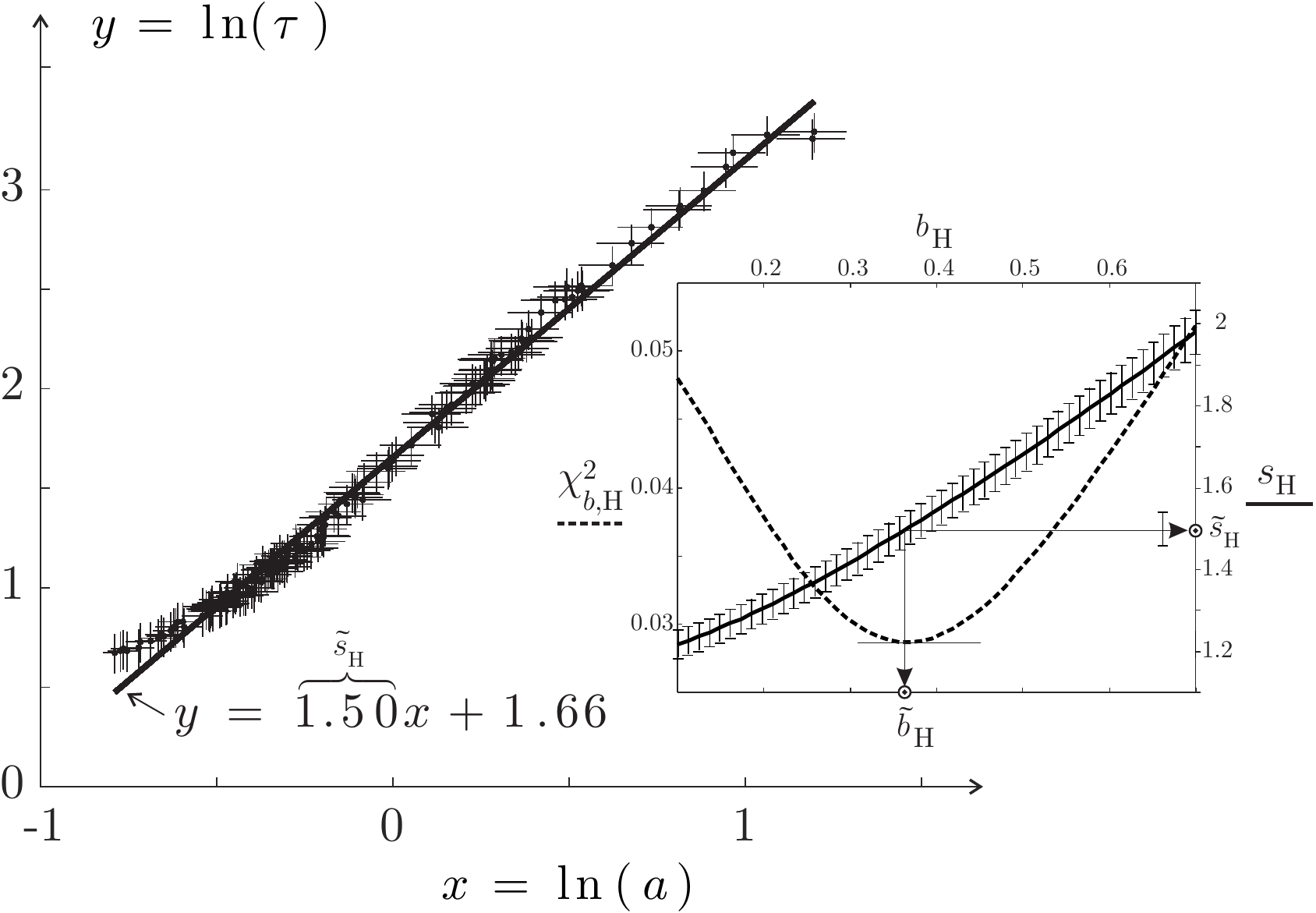}}
\caption{Isochrone test 1 for $\mathrm{X}=\mathrm{K}$ (upper panel) and $\mathrm{X}=\mathrm{H}$ (lower panel). In each panel the small box represents the residue $\chi_{b,\mathrm{X}}^2$ and the slope $s_{\mathrm{X}}$ of the weighted linear fit as functions of the isochrone parameter $b$. The residue plot is the dashed line with values on the left of the small box while the slope is the solid line with values on the right of the small box. The optimal value $\tilde{b}_{\mathrm{X}}$ of $b$ is identified as well as the corresponding optimal value of the slope $\tilde{s}_{\mathrm{X}}$ . The weighted best linear fit of the data $\left\{\ln(a),\ln(\tau)\right\}$ for $b=\tilde{b}_{\mathrm{X}}$ is then plotted.}
\label{piso}
\end{figure}

If each set of orbits is isochrone it must fulfill the generalized Kepler third law: $\tau^2\times a^{-3}=cst$ where $a$ is the isochrone length defined in section \ref{section2}. We then achieve an analysis in the space $\mathscr{H}_1=\left[\ln(a),\ln(\tau)\right]$. As the system is neither Keplerian, harmonic nor pseudo-H\'{e}non we guess it should be in a 
H\'{e}non potential; hence $a=\frac{1}{2}\left(\sqrt{r_a^2+b^2}+\sqrt{r_p^2+b^2}\right)$. In this formula $b$ is a macroscopic positive parameter common to all orbits, while $r_a$ and $r_p$ are microscopic ones, specific to each orbits. For a given value of $b$ we can plot the set $\left[\ln(a),\ln(\tau)\right]$ containing $n_\mathrm{X}$ points for each simulation $ \mathrm{X}=\mathrm{K}$ and $\mathrm{X}=\mathrm{H}$. We can then determine the weighted linear fit $y = s \ln(a)+ c$ (see appendix \ref{appendixfit}) of these plots and determine the residue of this fit namely 
\begin{equation}
\chi^2_{b,\mathrm{X}}=
\frac{1}
     {\ell^2} 
		\displaystyle\sum_{i=1}^{n_{\mathrm{X}}} \left\{ \ln(\tau_i)-\left[s\ln(a_i) +c \right]\right\}^2
		                       \;\mathrm{with}\;
		\ell^2= \displaystyle\sum_{i=1}^{n_{\mathrm{X}}} \ln(\tau_i)^2	.
\end{equation}
The optimal value $\tilde{b}_\mathrm{X}$ of $b_\mathrm{X}$ is obtained by a minimization algorithm applied to this residue computation. Explicitly, we compute the residue for discrete values of $b$ in the interval $\mathcal{B}=\left[0,b_{\max}\right]$ where $b_{\max}\simeq R_{50\%}$ as roughly estimated from the isochrone model. We then choose 
\begin{equation}
\tilde{b}_{\mathrm{X}}=\min _{b\in\mathcal{B}}\chi^2_{b,\mathrm{X}} .
\end{equation}
The plots of $\chi^2_{b,\mathrm{K}}$ and $\chi^2_{b,\mathrm{H}}$ as functions of $b$ are presented in the small boxes of figure \ref{piso}.
When the optimal value $\tilde{b}_{\mathrm{X}}$ is found, it should correspond to the optimal fit of $y=\ln(\tau)$ by a linear function of $x=\ln(a)$ for the considered simulation. The best slope $\tilde{s}_{\mathrm{X}}$ of this best linear fit corresponds to the best power law relation between $a$ and $\tau$, it should be 3/2 when the system is isochrone. We gather the statistics in table \ref{table:stat}. The uncertainty of $s_{\mathrm{X}}$ is evaluated by the dispersion parameter $\sigma_{b}$ defined in appendix \ref{appendixfit}. The amplitude of the error bar is equal to this dispersion parameter.
\begin{table}
\begin{tabular}
[c]{|l|c|c|c|c|}\hline
$\mathrm{X}%
\begin{array}
[c]{c}%
\\
\end{array}
$ & $\tilde{b}$ & $\chi_{b}^{2}\left(  \tilde{b}\right)  $ & $\tilde{s} $ & $\delta_{\tilde{s}}$\\ \hline
$\mathrm{H}%
\begin{array}
[c]{c}%
\\
\end{array}
$ & $3.70\times10^{-1}$ & $2.87\times10^{-2}$ & $1.50$ & $8.23\times10^{-2}%
$\\ \hline
$\mathrm{K}%
\begin{array}
[c]{c}%
\\
\end{array}
$ & $1.49\times10^{-1}$ & $1.07\times10^{-2}$ & $1.13$ & $4.06\times10^{-2}%
$\\ \hline
\end{tabular}
\caption{Statistical analysis for the sets $\left[\ln(a),\ln(\tau)\right]$ in the simulations $\mathrm{K}$ and $\mathrm{H}$.}
\label{table:stat}
\end{table}

\begin{figure}
\centerline{\includegraphics[width=\linewidth]{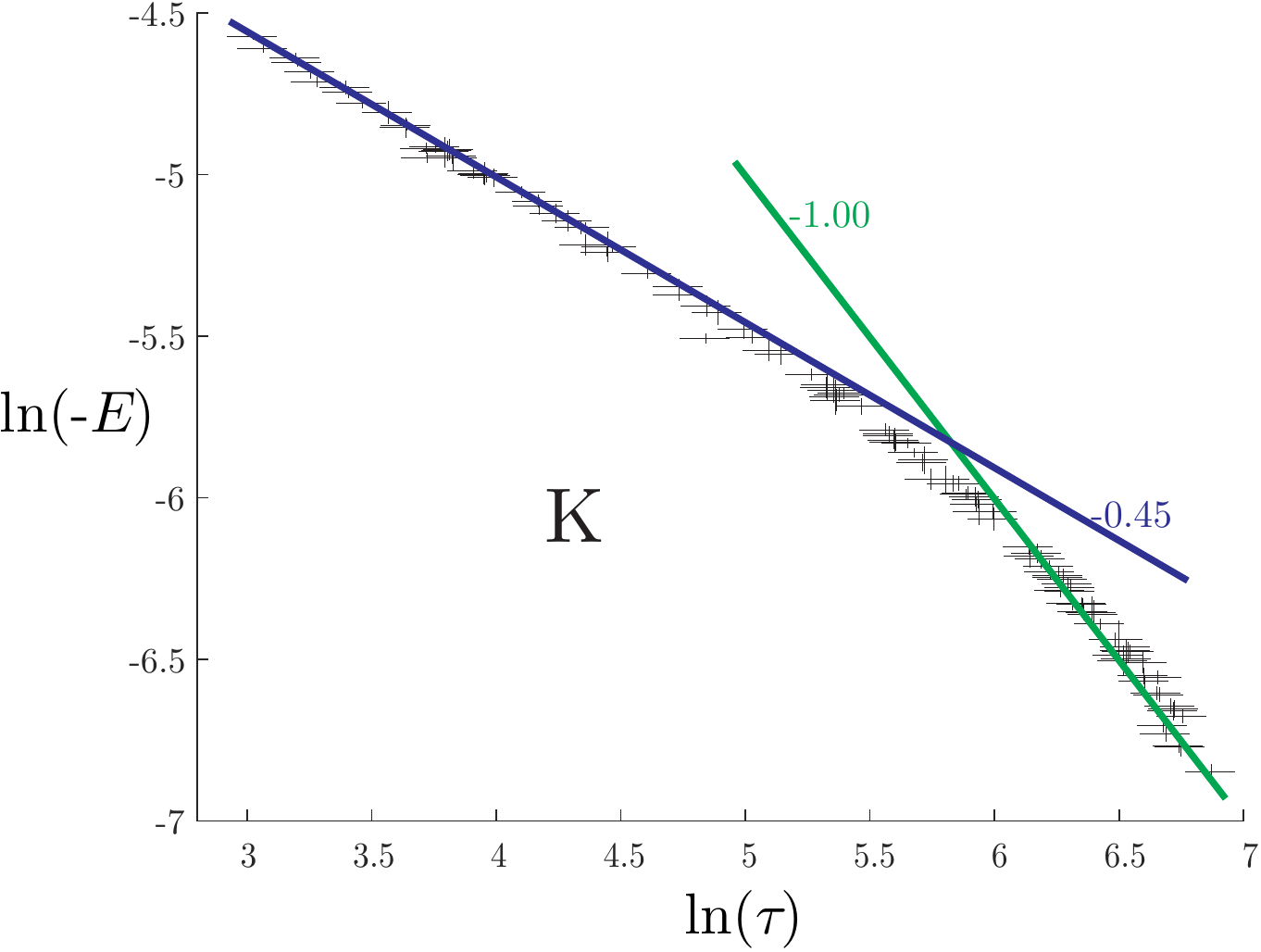}}

\centerline{\includegraphics[width=\linewidth]{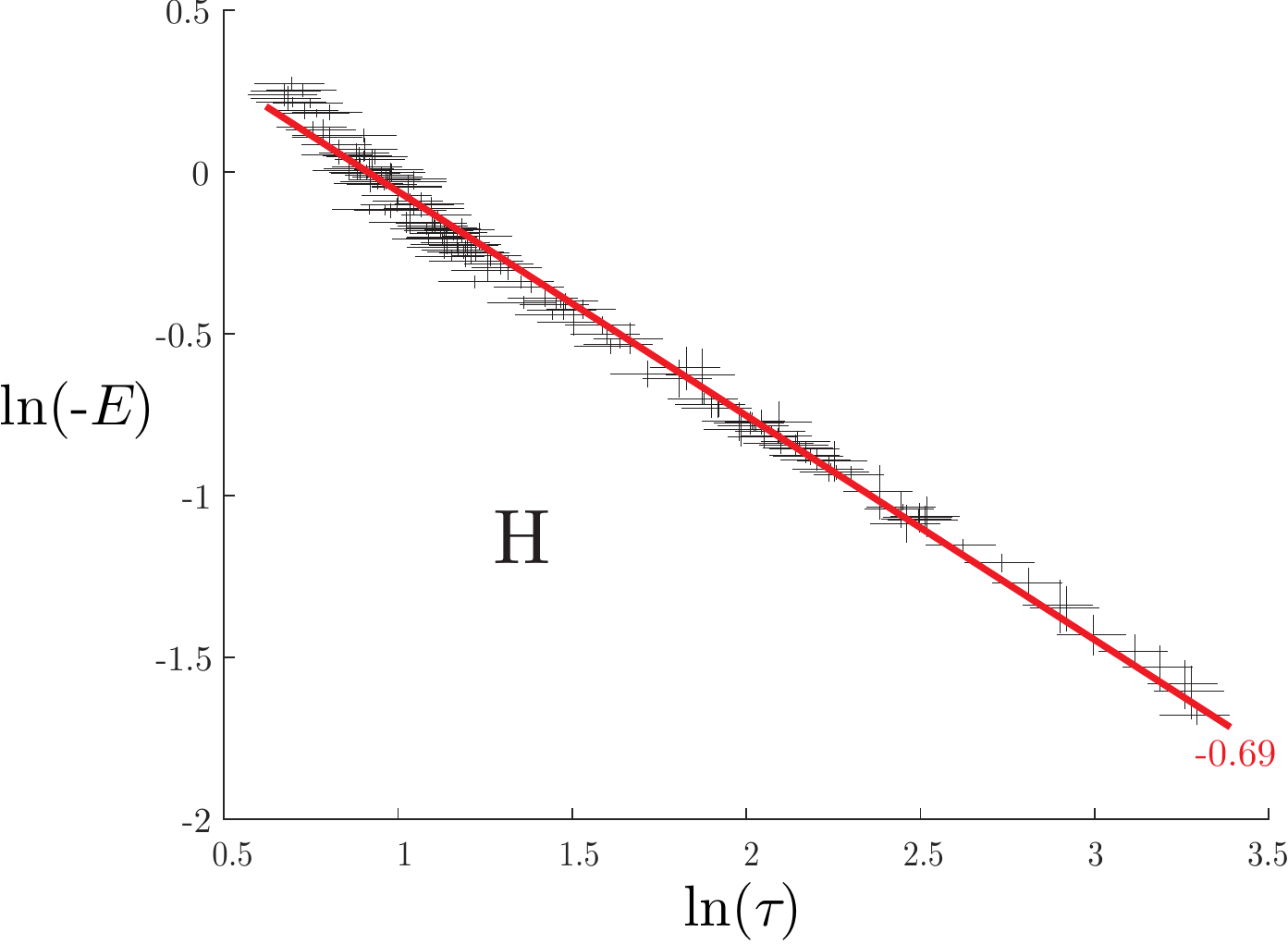}}
\caption{Isochrone test 2 for $\mathrm{X}=\mathrm{K}$ (top panel) and $\mathrm{X}=\mathrm{H}$ (bottom panel). The value of the slopes of the weighted linear fits are indicated near each line.}
\label{piso2}
\end{figure}

From our orbits analysis we can investigate another space, namely $\mathscr{H}_2=\left[\ln(\tau),\ln(-E)\right]$. This one is more direct than the previous because it does not need another parameter as $b$ to build the isochrone length $a$. For an isochrone model the Kepler third law applies and gives $\tau^2\times(-E)^3=cst$ which implies $\ln(-E)=-\frac{2}{3}\ln(\tau)+cst$. As we have computed the mean energy $E$ for each monitored orbit, we can additionally check if $\mathrm{K}$ or $\mathrm{H}$ should be isochrone in this sense. The results are plotted in figure \ref{piso2}.
\subsection{Analysis of the results}

The analysis of our results is unambiguous. Both spaces $\mathscr{H}_1$ and $\mathscr{H}_2$ reveal the isochrone nature of the result of the H simulation and the non isochrone nature of the K one. On the first hand the regression of $\ln(\tau)$ in $\ln(a)$ is perfectly linear for both simulations for the optimal value of $b$, but the value $\tilde{s}_{\mathrm{H}}=1,50\pm8,23. 10^{-2}$ is fully compatible with the isochrone one ${s}_{1,\mathrm{iso}}=3/2$ whereas $\tilde{s}_{\mathrm{K}}=1,13 \pm 4,063 . 10^{-2}$ is definitely not. 
On the other hand the analysis in $\mathscr{H}_2$ reveals that there is no unique power law relation between $\tau$ and $E$ for orbits in a $W_0 =9$ King model while this relation clearly exists with the right value ${s}_{2,\mathrm{iso}}^2=-2/3$ for the collapsed H\'{e}non sphere with an initial virial ratio $\kappa=-0.5$.

Why is the result of fast relaxation isochrone? The answer to this question was certainly proposed sixty years ago by Michel H\'{e}non in his seminal paper~\cite{Henon58}. During the mixing of the fast relaxation there is a natural tendency for particles to move toward equipartition in the energy. This is a pillar of statistical mechanics. In this context resonances enable energy exchanges between particles. If stars with the same radial period $\tau$ have different energies, they will exchange energy till they reach the same $E$. But the definition of isochrony is precisely that all stars with a given $\tau$ share the same $E$. 

\begin{figure}
\centerline{\includegraphics[width=0.8\linewidth]{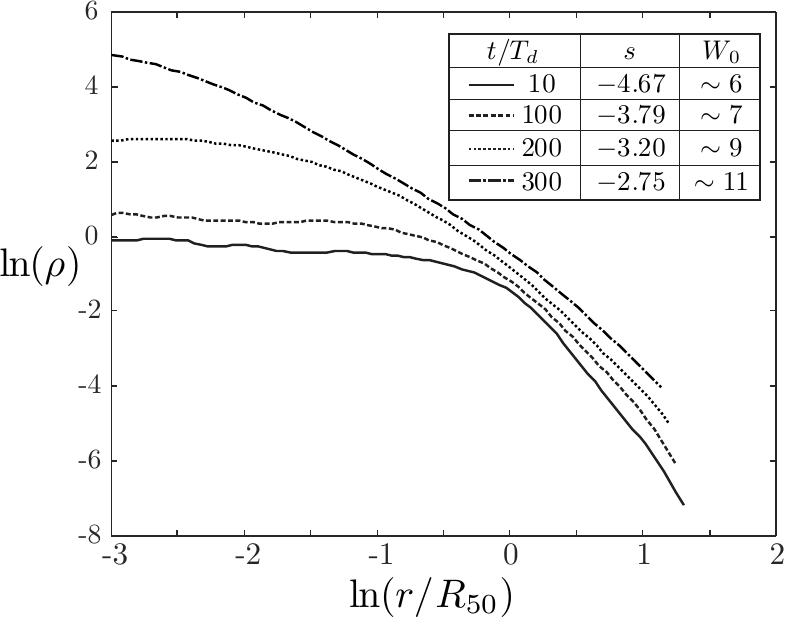}}
\caption{Evolution of the density for a H\'{e}non sphere with an initial virial ratio $\kappa=-0.5$ and a small softening parameter. The slope $s$ corresponds to the best linear fit of the halo. The value of $W_0$ is obtained by fitting the halo with a King model.}
\label{evoldens}
\end{figure}

Why has this old result been progressively forgotten? The response to this question is probably twofold. 
First of all, due to the slow relaxation process the initial state obtained after the fast relaxation is progressively changed into a more and more concentrated core-halo system. This gravothermal evolution was studied and understood during the last fifty years. During this process the mass density of the systems changes and it looses its isochrone property. When we let this possibility occur, decreasing the value the softening parameter by, at least, an order of magnitude ($\varepsilon\to\varepsilon/50$), the simulation of the same H\'{e}non sphere clearly shows this mass density evolution over the same duration (see figure \ref{evoldens}). Our statistical study  is no more possible in this context as the orbital periods evolve with the gravitational potential: when it is similar to a King model with $W_0=9$, it should have lost part of his isochrone property. 

Another simple reason of the isochrone oblivion is the efficiency of the King model: with 3 free parameters which control the slope of the halo, the size of the core  and the concentration of the system, this model is perfect to fit all the evolution of globular clusters evolution regarding the mass density or the luminosity profile. As long as nobody asks for detailed kinematical analysis there is no reason to discredit King model. 
   
\section{Conclusion and perspectives}

Let us summarize the main results we have presented in this paper:
\begin{itemize}
\item
The analysis of a self-gravitating system using only its mass density distribution (or luminosity profile) is ambiguous: several models can produce similar mass density profiles. In particular, the King model and its 3 free parameters can be used to produce a pretty good fit of the mass density profile of a globular cluster at any stage of its dynamical evolution.

\item 
The system produced just after the standard\footnote{By this precision we want to restrict this affirmation to systems which are initially not sufficiently cold to be influenced by the radial orbit instability.} fast relaxation process is a core-halo structure compatible with both a King or an isochrone mass density. However, when kinematic data is taken into account, the King model fails where the isochrone succeeds in reproducing the equilibrium state. 

\item
The isochrone model is just an initial condition obtained after the formation process of the system. Under slow relaxation processes the systems looses its isochrone character as it is confirmed by density profiles.
  
\end{itemize}

Such a result highlights the isochrone model in a new perspective. More than an aesthetic model, useful because it distinguishes itself by its ability for producing analytic formulas for the actions and angles of its orbits, the isochrone is in fact a fundamental potential resulting from a homogeneous fast relaxation process.

This process is the one supposed to occur in the formation of isolated self-gravitating systems (most globular clusters, LSB galaxies). It is then not surprising that they are characterized by a core-halo structure density profile as long they are not so affected by slow relaxation processes. When the formation process is hierarchical (e.g. HSB galaxies) the continuous merging process is combined with an inhomogeneous fast relaxation. The initial isochrone core-halo structure is no more observable as the core is unstable\footnote{In a inhomogeneous system the critical value of the density contrast could be very low. The collapse of the core could then appear during the merging and virializing phase}. This produces their cuspy profile. This could probably also explain the presence of supermassive black holes in the heart of such structures while they are not expected as a rule for globular clusters or LSB galaxies. 

Although the isochrone state is explicitly revealed after the homogeneous fast relaxation process, we have no more physical explanations than the one proposed by Michel H\'{e}non in the sixties (resonant coupling arguments). A special investigation on this subject is tricky because it requires to analyse orbits during a non-stationary phase. It deserves to be discussed in a future work.



\textbf{Acknowledgment} This work is supported by the ``IDI 2015" project
funded by the IDEX Paris-Saclay, ANR-11-IDEX-0003-02. The authors thank the referee
of the article for  
valuable comments and fruitful suggestions.



\bibliographystyle{plain}
\bibliography{Isochrony_GC_arxiv} 



\appendix

\section{Linear fit with weight }
\label{appendixfit}

Given a sequence $\left \{  x_{i};y_{i}\pm \sigma_{i}\right \}  _{i=1,\cdots,n}$, we have to compute values for
$c$ and $s$ which minimize the quantity
\begin{equation}
d^{2}=\sum_{i=1}^{n}\left[  y_{i}-\left(  c+sx_{i}\right)  \right]  ^{2}.%
\end{equation}
The uncertainty on $y_{i}$, namely $\sigma_{i}$, allows us to define the weight
$w_{i}=\sigma_{i}^{-2}$ of the $i-th$ value of $y$. The $d^2-$minimization problem
gives%
\begin{equation}
c=\frac{ 1 }{\Delta}\left(  \displaystyle \sum_{i=1}^{n}w_{i}x_{i}^{2}\right)  \left(\displaystyle \sum_{i=1}^{n}w_{i}y_{i}\right) 
    - \frac{ 1 }{\Delta}\left(\displaystyle \sum_{i=1}^{n}w_{i}x_{i}\right)  \left(
\displaystyle \sum_{i=1}^{n}w_{i}x_{i}y_{i}\right)
\end{equation}
and%
\begin{equation}
s=
 \frac{1  }{\Delta}\left( \displaystyle \sum_{i=1}^{n}w_{i}\right)  \left( \displaystyle \sum_{i=1}^{n}w_{i}x_{i}y_{i}\right)  
-\frac{1  }{\Delta}\left( \displaystyle \sum_{i=1}^{n}w_{i}x_{i}\right)  \left(
\displaystyle \sum_{i=1}^{n}w_{i}y_{i}\right)
\end{equation}
where
\begin{equation}
\Delta=\left(  \sum_{i=1}^{n}w_{i}\right)  \left(  \sum_{i=1}^{n}w_{i}%
x_{i}^{2}\right)  -\left(  \sum_{i=1}^{n}w_{i}x_{i}\right)  ^{2}.%
\end{equation}
The uncertainty on $c$ and $s$ are given by%
\begin{equation}
\sigma_{c}=\sqrt{\frac{1}{\Delta}\displaystyle \sum_{i=1}^{n}w_{i}x_{i}^{2}}\text{ and
}\sigma_{s}=\sqrt{\frac{1}{\Delta}\displaystyle \sum_{i=1}^{n}w_{i}}.%
\end{equation}


\end{document}